\documentclass[twocolumn]{aastex631}

\usepackage{mathptmx}
\usepackage[T1]{fontenc}
\usepackage{ae,aecompl}
\usepackage{graphicx}	
\graphicspath{{./Figures/}} 
\usepackage{amsmath}	
\usepackage{amssymb}	
\usepackage{subfigmat}
\usepackage{enumerate}
\usepackage{multirow}
\usepackage{natbib}
\usepackage{verbatim}

\newcommand{\rom}[1]{\uppercase\expandafter{\romannumeral#1}}
\newcommand{\Lsun}{$\text{L}_{\odot}$}

\accepted{December 5, 2023 in ApJ}

\shorttitle{Contemporaneous ejections and accretions in B335}
\shortauthors{C.-H. Kim et al.}

\begin{document}

\correspondingauthor{Jeong-Eun Lee}
\email{lee.jeongeun@snu.ac.kr}

\title{The CO outflow components ejected by a recent accretion event in B335}

\author[0000-0002-2523-3762]{Chul-Hwan, Kim}
\affil{Department of Physics and Astronomy, Seoul National University, 1 Gwanak-ro, Gwanak-gu, Seoul 08826, Korea}

\author[0000-0003-3119-2087]{Jeong-Eun Lee}
\affil{Department of Physics and Astronomy, Seoul National University, 1 Gwanak-ro, Gwanak-gu, Seoul 08826, Korea}
\affil{SNU Astronomy Research Center, Seoul National University, 1 Gwanak-ro, Gwanak-gu, Seoul 08826, Republic of Korea}

\author[0000-0003-1894-1880]{Carlos Contreras Pe{\~n}a}
\affiliation{Department of Physics and Astronomy, Seoul National University, 1 Gwanak-ro, Gwanak-gu, Seoul 08826, Korea}

\author[0000-0002-6773-459X]{Doug Johnstone}
\affiliation{NRC Herzberg Astronomy and Astrophysics,
5071 West Saanich Road,
Victoria, BC, V9E 2E7, Canada}
\affiliation{Department of Physics and Astronomy, University of Victoria,
3800 Finnerty Road, Elliot Building,
Victoria, BC, V8P 5C2, Canada}

\author{Gregory J. Herczeg}
\affil{Kavli Institute for Astronomy and Astrophysics, Peking University, Yiheyuan 5, Haidian Qu, 100871 Beijing, China}

\author{John J. Tobin}
\affiliation{National Radio Astronomy Observatory, 520 Edgemont Rd., Charlottesville, VA 22903}

\author{Neal J. Evans II}
\affiliation{Department of Astronomy, The University of Texas at Austin, 2515 Speedway, Austin, TX 78712, USA}

\begin{abstract}
Protostellar outflows often present a knotty appearance, providing evidence of sporadic accretion in stellar mass growth.
To understand the direct relation between mass accretion and ejection, we analyze the contemporaneous accretion activity and associated ejection components in B335. B335 has brightened in the mid-IR by 2.5 mag since 2010, indicating increased luminosity, presumably due to increased mass accretion rate onto the protostar. ALMA observations of $^{12}$CO emission in the outflow reveal high-velocity emission, estimated to have been ejected 4.6--2 years before the ALMA observation and consistent with the jump in mid-IR brightness.
The consistency in timing suggests that the detected high-velocity ejection components are directly linked to the most recent accretion activity. We calculated the kinetic energy, momentum, and force for the ejection component associated with the most recent accretion activity and found that at least, about 1.0$\%$ of accreted mass has been ejected. More accurate information on the jet inclination and the temperature of the ejected gas components will better constrain the ejected mass induced by the recently enhanced accretion event.

\end{abstract}


\section{Introduction} \label{sec:introduction}
Episodic accretion plays an important role in stellar mass assembly, including as one of the solutions to the luminosity problem \citep[e.g.,][]{Offner2011, dunham2012, Fischer2017}. Theoretical models and observational evidence both suggest that episodic accretion is more extreme and frequent in the Class0/\rom{1} stages \citep[e.g.,][]{Vorobyov2015, park2021, zakri2022}. Although a few protostellar outbursts have been observed via brightness variations that last for years or even decades \citep{kospal2007, Safron2015, park2021}, most observational evidence for episodic accretion comes from knotty features in the outflow/jet \citep{Hirano2006, Lee_CF2009, Lee_JE2010, plunkett2015nat}, or chemical signatures in envelopes \citep[e.g.,][]{jelee07, kimhj2012, Jorgensen2015}.\par
Outflows have been ubiquitously found around young stellar objects (YSOs) \citep[e.g., ][]{Hatchell2009} since the first detection 40 years ago \citep{Snell1980}. Two types of outflows are commonly observed: (1) wide-opening slow (few to 10 km s$^{-1}$) molecular outflows, typically observed with molecular lines such as CO, and (2) well-collimated fast (100 km s$^{-1}$) outflows/jets, typically observed in the near-infrared with forbidden iron emission [Fe \rom{2}] and H$_{2}$ or seen in shock-tracing molecular species such as CO and SiO \citep{Bally2016, bjerkeli2019, Dutta2022Early, cfLee2020}.\par
In addition, well-collimated outflows/jets consist of gas launched from the YSO and often present a knotty appearance \citep[e.g.][]{Hirano2006, Lee_CF2009, Lee_JE2010, plunkett2015nat}, indicative of variable, potentially discrete, ejection events presumably linked to an unsteady and possibly eruptive mass accretion process. 
\citet{Dutta2022Early, Dutta2022Evolved} reported the knot features of SiO and CO emission observed in the protostellar stages from early to evolved and calculated the jet mass-loss rate. \citet{Jhan2022} also showed the presence of knotty features of SiO emission in six Class 0 and \uppercase\expandafter{\romannumeral1} sources.
\par

The accretion-related YSO variability is another observational evidence of episodic accretion. While historical searches for outbursts focused on optical wavelengths, YSOs at the Class 0/\rom{1} stages are deeply embedded in the surrounding envelope, thus YSOs at these stages cannot be observed in the optical/near-infrared. However, sustained mid-infrared monitoring by \textit{WISE/NEOWISE} \citep{carlos2020, park2021} and sub-mm monitoring at JCMT \citep{lee_yh2021} revealed the variability of YSOs at these stages. \citet{park2021} performed an ensemble study to classify YSO variability using 6.5 years (14 epochs) of \textit{NEOWISE} light curves for 717 Class 0/\rom{1} protostars, identified by previous Spitzer and Herschel surveys \citep{Megeath2012, Dunham2015, Esplin2019}. They found that $\sim55\%$ of Class 0/\uppercase\expandafter{\romannumeral1} protostars can be classified as variable stars, with the majority undergoing small amplitude - order unity - variations and a smaller fraction having much larger brightness changes. Furthermore, the mid-IR variability of protostars has been shown to be typically a reliable tracer of accretion luminosity changes in these systems \citep{carlos2020}.\par

Then, the protostellar brightness variability should be causally connected with the knotty outflow/jet features; the mass flow onto the forming star is intimately linked to accretion through the disk and angular momentum dispersal from the disk via an outflow in an unstable process. However, only tentative connections have been identified between an increase in protostellar outflow activity and a known luminosity burst \citep[e.g.,][]{Ellerbroek2014}. 
Definitive measurements of an outflow event coupled to a specific accretion burst would demonstrate an  incontrovertible physical link. In this paper, we present a unique case study of B335, where the {\it contemporaneous} accretion and ejection events were revealed.

B335 is an isolated dark globule located at a distance of 164.5 pc \citep{Watson2020}. 
Previous studies \citep{yen2010, evans2015, Yen2015, yen2015_b335} revealed infalling motions toward the center of B335, while no clear rotation motions are observed at large scales. The lack of clear rotation motion at 1000\,AU implies that the size of the protoplanetary Keplerian disk is $<$16 AU if it exists \citep[using the updated distance of 164.5\, pc;][]{yen2010, Yen2015, yen2015_b335}. \citet{Imai2016, imai2019} have reported a number of complex organic molecules (COMs) from B335, suggesting that B335 harbors a hot corino associated with the central Class 0 protostar IRAS 19345+0727. \par

The large-extended outflow of $^{12}$CO $J=1-0$ for B335 was revealed by \citet{hirano1988}, nearly 2\arcmin\ (north to south) $\times$ 8\arcmin\ (east to west). \citet{yen2010} reported the high-velocity $^{12}$CO $J=2-1$ component in B335 that have size of 6\arcsec\ (north to south) $\times$ 10\arcsec\ (east to west).
Recently, \citet{bjerkeli2019} used $^{12}$CO $J=2-1$ of combined high spatial resolution ALMA data to uncover a hint of a recent accretion outburst as a high-velocity ejection component in the B335. However, direct evidence of the accretion event responsible for the ejection activity was absent from that analysis. Therefore, here we analyze the \textit{WISE/NEOWISE} light curve of B335 to study its \textit{time-dependent} variability and connect it with the fossil record of the \textit{time-dependent} evolution of the outflow/jet observed with ALMA for the same high-velocity ejection component.

In this work, we aim to investigate \textit{the relationship between contemporaneous accretion and ejection activities} in B335. Matching the properties of the ejecta with the recent accretion history greatly aids the determination of the mass accreted onto the protostar versus the mass dispersed away from the system.
We present the observations and data reduction in Section \ref{sec:Observation}. Our results are described in Section \ref{sec:Analysis}. We discuss the physical relationship between the derived contemporaneous mass accretion rate and mass ejection rate in Section \ref{sec:comparision_eje_acc}. The source of uncertainty in ejection mass estimation is discussed in Section \ref{sec:uncertainty}. Finally, conclusions are given in Section \ref{sec:Conclusions}.

\begin{deluxetable*}{cccccc}[!htp]
\tabletypesize{\scriptsize}
\tablecolumns{5}
\tablecaption{Summary of ALMA Band 6 data used in this study\label{tab:summary_obs_table}}
\tablenum{1}
\tablewidth{0pt}
\tablehead{\colhead{Project ID} & \colhead{Date} & \colhead{Continuum/Cube data} & \colhead{Beam size} & \colhead{RMS noise level}\\
           \colhead{} & \colhead{} & \colhead{} & \colhead{(\arcsec)} & \colhead{(mJy beam$^{-1}$)}}

\startdata
2017.1.00288.S & 2017.10.21-29 & Continuum & 0.046 $\times$ 0.044   & $\sim$0.015\tablenotemark{a}\\
               &            & $^{12}$CO & 0.092 $\times$ 0.081   & $\sim$1.70\tablenotemark{b} \\
\enddata
\tablenotetext{a}{The RMS noise level of continuum image was estimated from the emission-free region.}
\tablenotetext{b}{The RMS noise level of the $^{12}$CO maps was calculated per channel (channel width = 0.16 km s$^{-1}$) and averaged over the line-free channels.}
\end{deluxetable*}
\vspace{-9mm}

\section{Observational Data} \label{sec:Observation}
\subsection{ALMA} \label{subsec:ALMA}
The ALMA data archive contains several observational datasets of B335\footnote{\protect\url{https://almascience.nao.ac.jp/aq/?result_view=observation&sourceName=B335}}. For the analysis presented in this work we obtained the highest-angular resolution data available in the archive. A high-resolution observation was carried out with ALMA in band 6 between October 21 and 29, 2017 (2017.1.00288.S PI: Bjerkeli, Per). The data contained five execution blocks and cover five spectral windows (SPWs). The SPWs included the $^{12}$CO $J = 2-1$ transition, $^{13}$CO $J = 2-1$ transition, C$^{18}$O $J = 2-1$ transition, SiO $J = 5-4$ transition, and continuum emission, respectively. In particular, the spectral resolution of the data, including $^{12}$CO $v = 0$ $J = 2-1$ (230.538 GHz), was set to 122.070 kHz (0.16 km s$^{-1}$, $\Delta \textit{v}$) while the total bandwidth was 117.187 MHz. The continuum SPW was set to the spectral resolution of 15.625 MHz and a total bandwidth of 2.000 GHz. The data were initially calibrated using CASA v5.1.1 \citep{McMullin2007}. 

To obtain a better signal-to-noise ratio (SNR), we performed self-calibration for the continuum emission using CASA v6.2.1 \citep{McMullin2007}. The self-calibrated continuum was imaged with natural weighting using the same CASA version of the CLEAN algorithm. The data cube with $^{12}$CO had the continuum emission self-calibration solution applied and was then imaged from the continuum subtracted visibility data using \textit{tclean} task at CASA v6.2.1. For the CO image, we tested three different weightings: uniform, natural, and Briggs with a robust parameter of 0.5. The high-velocity ejection emission was not detected in the uniform weighting image, and the SNR of the emission in the Briggs weighting image was too low. As a result, we adopted a natural weighting and a uvtaper of 0.06\arcsec\ to improve the SNR.

\begin{figure}[htp]
\centering
{
\includegraphics[width=1\linewidth]{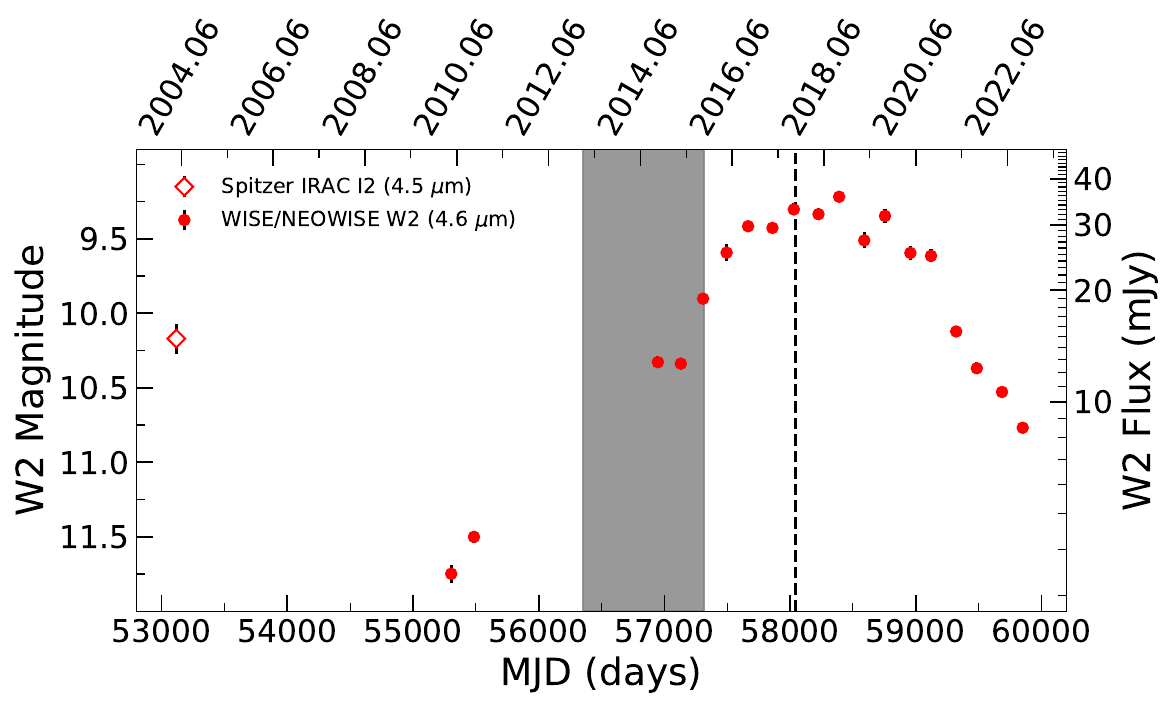}
}
\caption{Light curve of B335 in the IRAC I2 band of \textit{Spitzer}, W2 band of \textit{WISE}, \textit{NEOWISE}-Post Cryogenic, and \textit{NEOWISE} survey. The short solid lines attached to the data points represent the magnitude uncertainties at given epochs. The black dashed vertical line indicates the ALMA observation date (October 2017). The gray-shaded region represents the period when the high-velocity ejection clumps identified in the PV diagram shown in Figure 2 were ejected (March 2013 - October 2015).
\label{fig:b335_w2_lc}
}
\end{figure}

\begin{figure*}[htp]
\centering
{
{
\includegraphics[width=1\columnwidth,keepaspectratio]{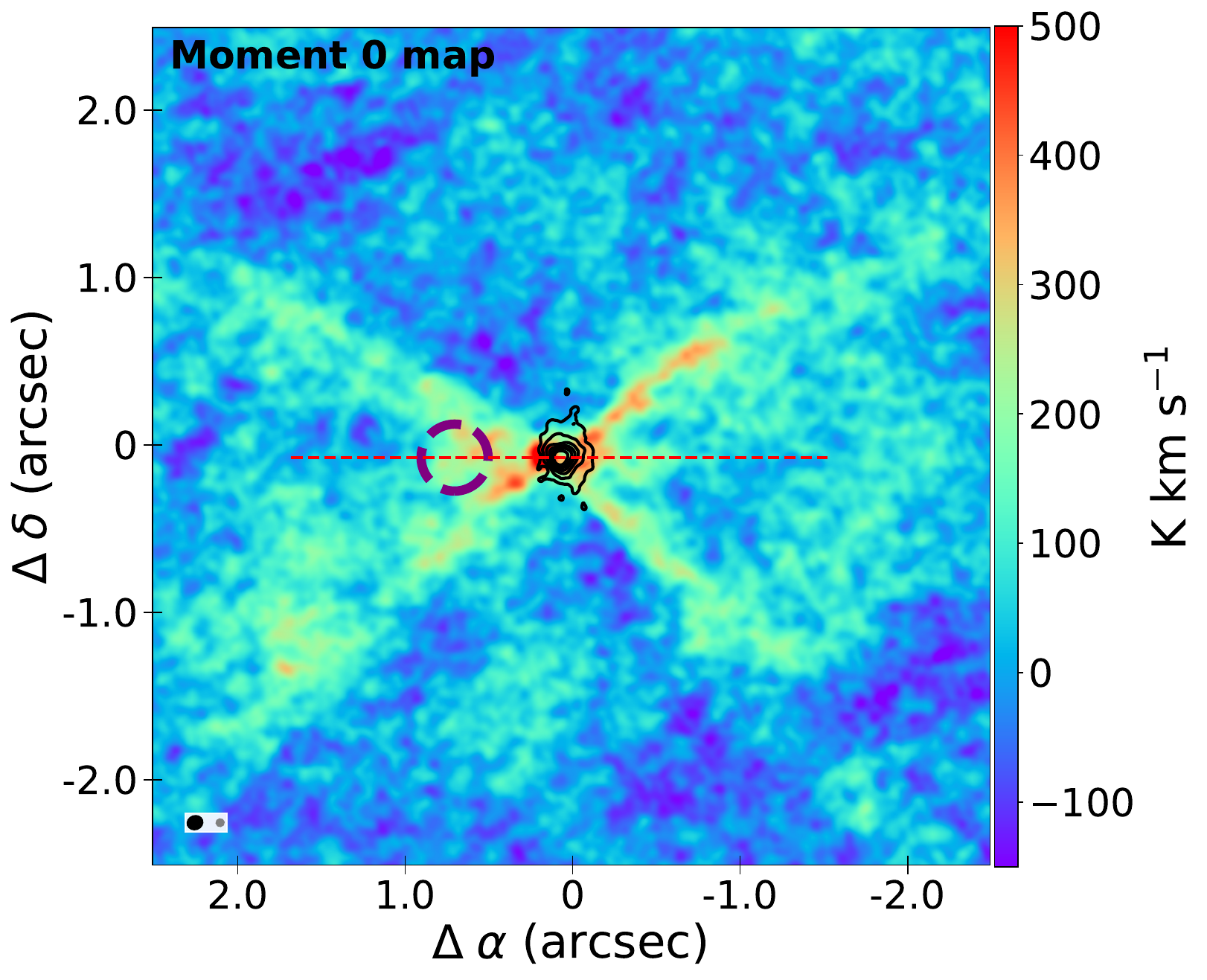}
\includegraphics[width=0.9\columnwidth,keepaspectratio]{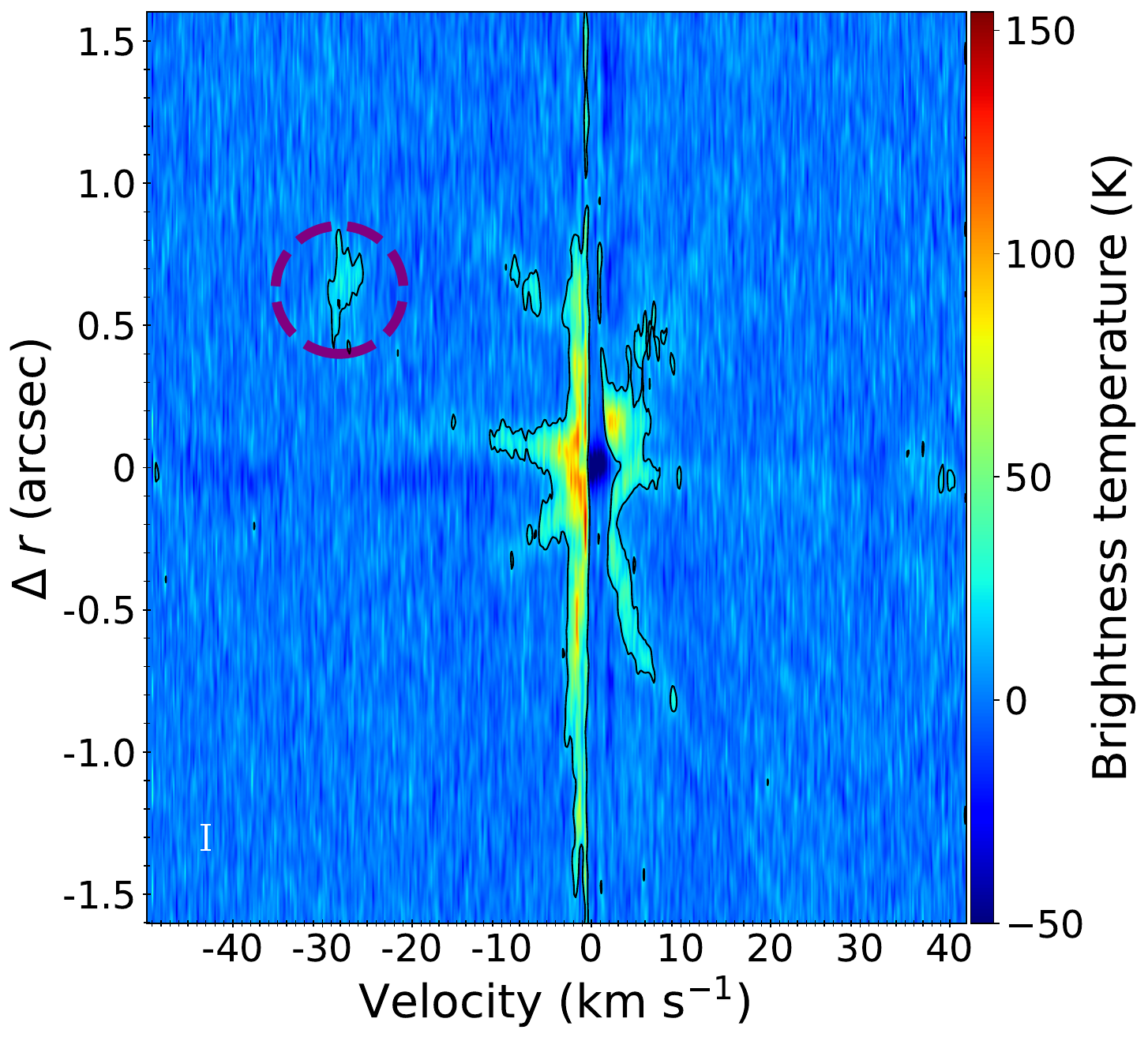}
}
}
\caption{\textit{Left}: Integrated intensity map of $^{12}$CO over the velocity range (-49.6 km s$^{-1}$ < \textit{v$_{\text{range}}$} < 41.7 km s$^{-1}$).
Black contours represent continuum emission. Contour levels are from 10, 20, 40, 80, 160, 320 $\sigma_{\text{cont}}$ ($\sigma_{\text{cont}}$ = 0.015 mJy beam$^{-1}$). 
The synthesized beams for the $^{12}$CO cube data and continuum data are presented at the lower left corner of the map as a black ellipse and gray ellipse, respectively.
The red dashed line represents the direction of the cut for a PV diagram. The purple dashed circle represents the location of the high-velocity ejection component indicated in the right panel. \textit{Right}: PV diagram for $^{12}$CO along the direction of the red dashed line shown in the left panel. Black contours indicate signals higher than 3 $\sigma$ ($\sigma$ = 1.7 mJy beam$^{-1}$ $\simeq$ 5 K).
The high-velocity ejection components are marked with a purple dashed circle. 
The FWHM of the synthesized beam ($\sim$0.087\arcsec) and the velocity resolution (0.16 km s$^{-1}$) are indicated with the vertical and horizontal bars, respectively. We note that the mark for the velocity resolution is too small to be visible.
\label{fig:b335_mom0_pv}
}
\end{figure*}

\subsection{WISE/NEOWISE} \label{subsec:WISE}
The \textit{Wide-field Infrared Survey Explorer} (\textit{WISE}) is a mid-IR survey of the entire sky using four bands, W1 (3.4 $\mu$m), W2 (4.6 $\mu$m), W3 (12 $\mu$m), and W4 (22 $\mu$m), with an angular resolution of 6.1\arcsec, 6.4\arcsec, 6.5\arcsec, and 12.0\arcsec, respectively, from January to September 2010 \citep{wright2010}. The \textit{WISE} mission continued to operate for an additional four months using W1 and W2 bands  after the depletion of hydrogen for the cryostat of \textit{WISE}. This was known as the \textit{NEOWISE} Post-Cryogenic Mission \citep{mainzer2011}. In September 2013, \textit{WISE} was reactivated as the \textit{NEOWISE} mission and has been performing observations in the W1 and W2 bands with a cadence of six months since 2013 \citep{mainzer2014}. In each visit to an area of the sky, {\it WISE} performed 10-20 observations over a period of 1-3 days. 

B335 has been observed by the \textit{WISE}/\textit{NEOWISE} survey since 2010: in April and October 2010 and every six months since 2014. To construct the source light curve, single-epoch catalogs found at the NASA/IPAC Infrared Science Archive were queried using a 5\arcsec\ radius around the coordinates of B335. The W2 band data used in this work was averaged applying a similar method from \citet{park2021}. The choice in radius was based on previous works using \textit{WISE}/\textit{NEOWISE}, especially in the analysis of protostars \citep[see ][]{carlos2020}. The method of \citealp{park2021} estimates a mean right ascension and declination, or (mRA, mDEC) from all of the points selected within the seach radius. Then for every point, it estimates the distance to the (mRA, mDEC) coordinates. Finally, the average photometry is derived only from points that are within 2$\sigma$ in distance from (mRa, mDEC). Single-epoch images were also visually inspected to ensure that the photometry provided accurate results. For a more detailed description of how the single-epoch data is averaged, see \citet{park2021}.


\section{Analysis} \label{sec:Analysis}

\subsection{The mid-IR outburst} 
\label{sec:lc_b335}
Figure \ref{fig:b335_w2_lc} presents the 4.6 $\mu$m mid-IR light curve of B335. The data includes the \textit{WISE}/\textit{NEOWISE} W2 data described in Section \ref{subsec:WISE}, as well as Spitzer IRAC I2 band photometry. The latter value is derived by performing aperture photometry on mosaic images obtained from the NASA/IPAC Infrared Science Archive (IRSA). The aperture size was set at 12 \arcsec~following the guidelines from the IRAC instrument handbook. Finally, the IRAC I2 photometric point was corrected to match the WISE W2 passband following the equations of \citet{2014Antoniucci}.

The figure shows that B335 brightened by 2.5 magnitudes between \textit{WISE} and \textit{NEOWISE} observations, reaching a peak brightness of W2$=9.2$ mag. Due to the gap between \textit{WISE} (55304 MJD) and the start of \textit{NEOWISE} (56949 MJD) observations, it is difficult to interpret the shape of the outburst. The light curve appears to show that YSO brightened smoothly between 2010 and 2017-2018 when it reached its peak brightness, but a quiescent phase followed by a sudden outburst around 2013-2014 could also describe the shape of the light curve. The additional {\it Spitzer} point also adds some uncertainty in interpreting the light curve. The brighter photometric point could be interpreted as B335 going through periodic outbursts. Such behavior, which could result from dynamical perturbations from stellar or planetary companions, has been observed in an increasing number of YSOs \citep{2015Hodapp, 2020Dahm, 2022Guo}.

In spite of the difficulties in interpreting the shape of the outburst, the high-amplitude variability observed between \textit{WISE} and \textit{NEOWISE} observations is likely associated with a recent accretion outburst of B335.
The ALMA observation was taken in October 2017 when the outburst in the YSO had already reached its maximum brightness.

\begin{figure*}[!htb]
\centering
{
\includegraphics[width=1\columnwidth,keepaspectratio]{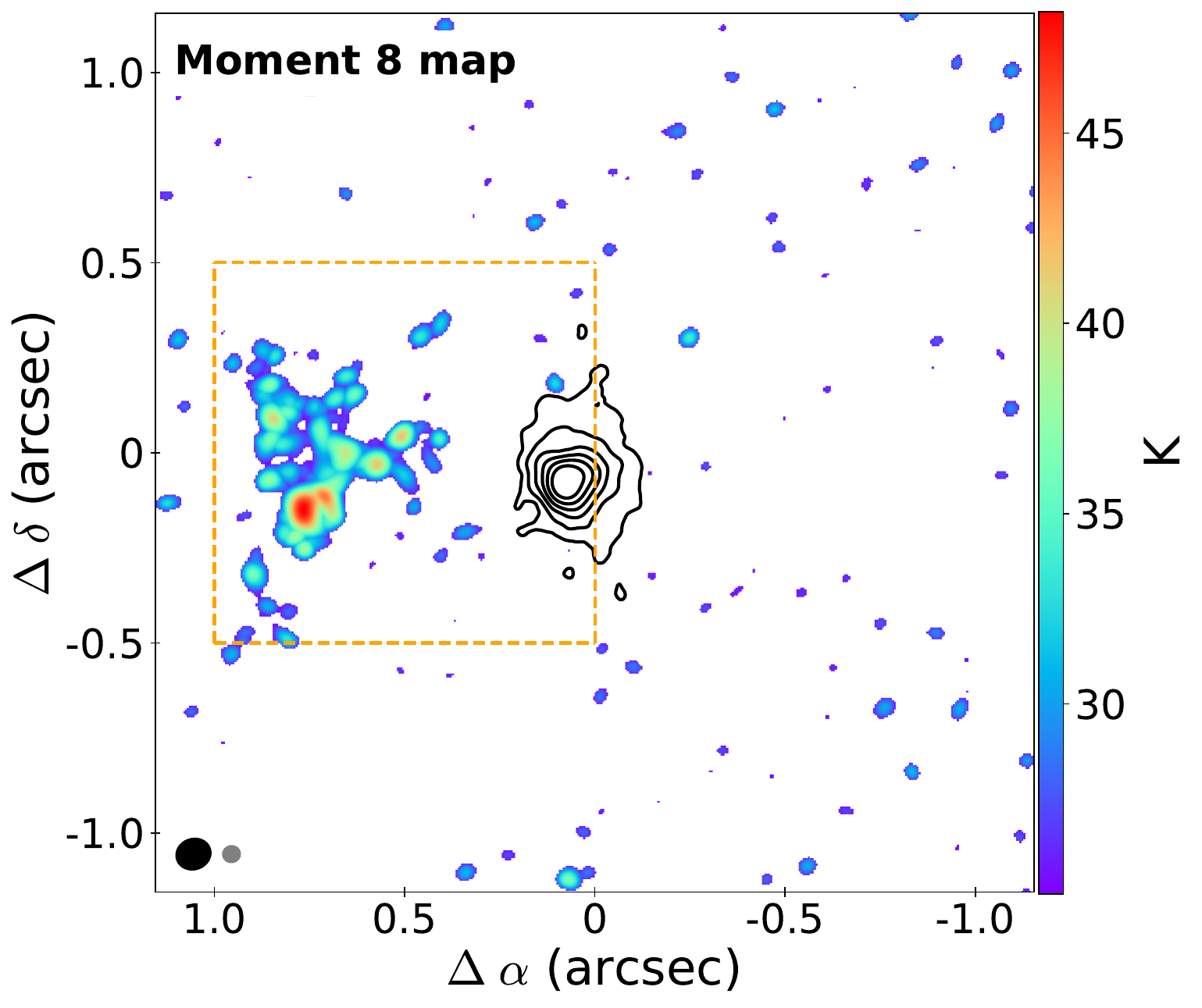}
\includegraphics[width=1.03\columnwidth,keepaspectratio]{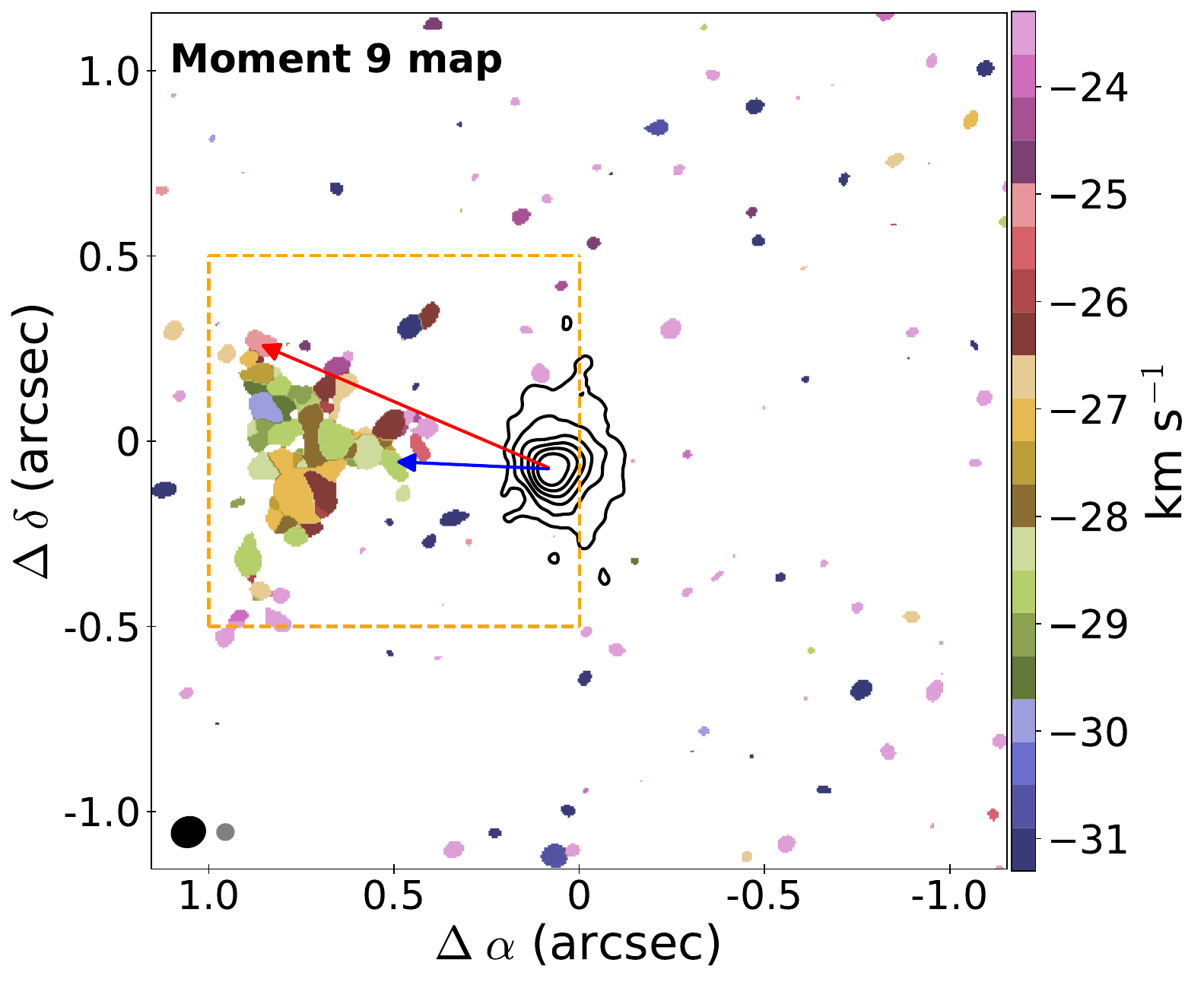}
}
\caption{\textit{Left}: Peak intensity (moment 8) map of the high-velocity $^{12}$CO clumps. \textit{Right}: Velocity of the peak intensity (moment 9) map of the high-velocity $^{12}$CO clumps. The velocity range used for both maps is from $-37.3$ km s$^{-1}$ to $-17.3$ km s$^{-1}$, and only emission above 5 $\sigma$ was used. The long red and short blue arrows indicate the distances from the central continuum peak to the earliest and latest ejected clumps among the high-velocity $^{12}$CO clumps. The black contours represent continuum emission. Contour levels are from 10, 20, 40, 80, 160, 320 $\sigma_{cont}$. The synthesized beams for each moment map and the continuum data are presented at the lower left corner of the maps as black and gray ellipses, respectively. The orange dashed box indicates the region over which the total mass of the high-velocity ejection components is calculated.}
\label{fig:b335_mom8_9_pv}
\end{figure*}

\subsection{The high-velocity $^{12}$CO clumps ejected by the recent outburst}
\label{sec:comparison_lc_clumps}
B335 is estimated to have a systemic velocity of 8.3--8.5 km s$^{-1}$ \citep[$\textit{v}_{sys}=8.3$\,km\,s$^{-1}$ adopted here,][]{evans2005, jorgensen2007, yen2011, mottram2014}. The systemic velocity was taken into account for all velocities used in this paper. The outflow inclination is $\sim$10$^\circ$ from the plane of the sky \citep{hirano1988}. \citet{stutz2008} reported the inclination of the disk of 87$^\circ$ from the line of sight (nearly edge-on), corresponding to the outflow inclination of 3$^\circ$ from the plane of the sky. We adopt 10$^\circ$ as the standard to derive outflow properties but explore the effect of the smaller inclination in Section \ref{subsec:Inclination}.

Figure \ref{fig:b335_mom0_pv} presents the integrated intensity map (left) and the Position-Velocity (PV) diagram of the $^{12}$CO 2--1 line made along the outflow axis (right). The PV diagram reveals the high-velocity ejection component at $-27.3$ km s$^{-1}$. Although the SiO and $^{13}$CO emission were not detected at the high-velocity, the same CO high-velocity component is also found in the analysis by \citet{bjerkeli2019}. Figure \ref{fig:b335_mom8_9_pv} shows the peak intensity map and the velocity map at the peak intensity, where the considered velocity range is $-37.3$ km s$^{-1}$ to $-17.3$ km s$^{-1}$, which fully covers the high-velocity ejection component.
It is also revealed that this high-velocity ejection component consists of multiple components in Figure \ref{fig:b335_mom8_9_pv}.

Outflows commonly show a bipolar structure. We thus examined whether red-shifted high-velocity ejection components associated with the blue-shifted high-velocity ejection components could be found and confirmed that no red-shifted high-velocity ejection components were detected in the ALMA observations.

To calculate the time when the high-velocity components were ejected, their velocities and distances between these components and the protostar are needed, taking into account the inclination of the target. The earliest ejected high-velocity component has a velocity of 145.6 km s$^{-1}$, after adjusting for the 10$^\circ$ inclination of the outflow. It is located at a distance of $\sim$144.4 au from the central source.  The most recent ejected component has a velocity of 165.7 km s$^{-1}$ and a distance of $\sim$71.0 au (see the red and blue arrows, respectively, in the right panel of Figure \ref{fig:b335_mom8_9_pv}).
We estimate the kinematic timescale of both components by dividing the distance by the velocity. Therefore, we infer that the high-velocity ejection components were ejected 4.6 to 2.0 years before the ALMA observation date. The gray-shaded region in Figure \ref{fig:b335_w2_lc} represents the timescale over which the high-velocity ejection clumps were ejected. This is consistent with the period of increase when B335 brightened significantly. 

The outflow inclination of 10$^{\circ}$ was provided by \citet{hirano1988} without a measurement error, making it challenging to estimate the uncertainty in the ejection period of high-velocity clumps. For an estimation, we assume a 30\%\ uncertainty in the outflow inclination, i.e., 10$\pm$3$^{\circ}$ and adopt the velocities and distances for the earliest and latest high-velocity ejection components. The ejection period ranges from 6.0 to 2.6 yr before the ALMA observation date for the outflow inclination of 13$^{\circ}$ and from 3.3 to 1.4 yr for the outflow inclination of 7$^{\circ}$. As a result, the uncertainty of the ejection period could be up to 3 years given the 30\% uncertainty of the outflow inclination.

\section{Comparision between Ejection and Accretion} \label{sec:comparision_eje_acc}
\subsection{Properties of ejected gas components} \label{sec:dis_properties}
We compute the properties (mass, $M_{\text{ejection}}$; kinematic energy, $E_{\text{ejection}}$; momentum, $P_{\text{ejection}}$; and force, $F_{\text{ejection}}$) of the high-velocity ejection clumps following \citet{dunham2014}.
Since the ejection properties depend on the mass and the H$_{2}$ column density is required to estimate the mass of ejected gas,
we first calculate $N(\text{H}_{2})_{\textit{v}, \text{pixel}}$ using the $^{12}$CO data for each pixel and each velocity channel within a selected range, where we assume that the ejection components are optically thin and follow local thermal equilibrium (LTE) conditions. The column density of the H$_{2}$ then can be estimated using:

\begin{align}\label{eq:columndensity_h2}
    N(\text{H}_{2}) &= f(J, T_{\text{ex}}, X_{\text{mol}})T_{\text{mb}}\Delta \textit{v},
\end{align}
\begin{align}\label{eq:func_quantum}
    f(J, T_{\text{ex}}, X_{\text{mol}}) &= X_{\text{mol}}^{-1}\frac{3k}{8\pi^{3}\nu\mu^{2}}\frac{(2J+1)}{(J+1)}\frac{Q(T_{\text{ex}})}{g_{J}}e^{\frac{E_{J+1}}{kT_{\text{ex}}}} ,
\end{align}
where $f(J, T_{\text{ex}}, X_{\text{mol}})$ is a function of the quantum number of the lower rotation state, $J$, the molecule abundance relative to H$_{2}$, $X_{\text{mole}}$, and the excitation temperature of the ejected material, $T_{\text{ex}}$. $T_{\text{mb}} \Delta \textit{v}$ is the temperature integrated for each pixel and velocity. 

In Equation \ref{eq:func_quantum}, we adopt a standard CO abundance relative to H$_{2}$, $X_{\text{CO}}=1.7\times10^{-4}$ \citep{Lacy2017}. The rest frequency of $^{12}$CO $J = 2-1$ corresponds to $\nu = 230.538$ GHz, the $^{12}$CO dipole moment has a value of $\mu = 0.11$ Debye, and the upper energy level, $E_{J+1}$, equals  $2.291\times10^{-15}$ erg. The partition function, $Q(T_{\text{ex}})$, and the degeneracy of the lower state, $g_{J}$, are expressed as $\sum_{J=0}^{\infty} g_{J}e^{-E_{J}/kT_{\text{ex}}}$ and $2J+1$, respectively. 

The excitation temperature, $T_{\text{ex}}$, is an unknown parameter when calculating $N(\text{H}_{2})$. Previous studies have found evidence for warm gas in outflows \citep[e.g.][]{hatchell1999a, hatchell1999b, nisini2000, giannini2001}. For example, \citet{hatchell2007} adopted a value of $T_{\text{ex}}=50$ K by considering the evidence of warm gas in outflow, while others \citep{green2013, jeLee2014, Je2015} found a higher $T_{\text{ex}}$, close to 350 K, for the warm CO gas using higher \textit{J} levels. 

\begin{figure}[htp]
\centering
{
\includegraphics[width=1\linewidth]{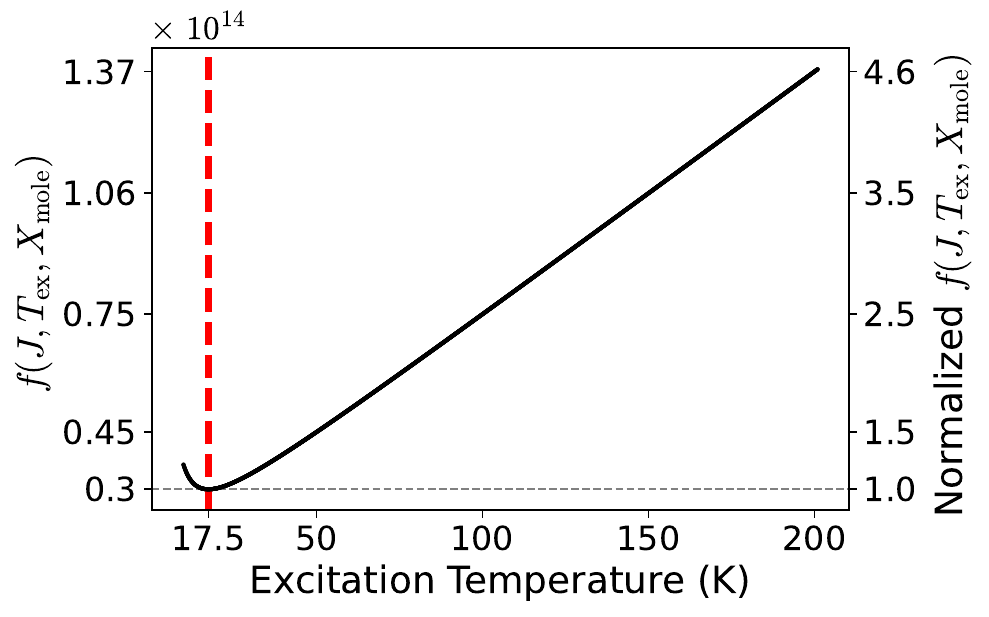}
}
\caption{\textit{Left y-axis: }value of $f(J, T_{\text{ex}}, X_{\text{mol}})$ as a function of $T_{\text{ex}}$ for the $^{12}$CO $J = 2-1$ transition line. The red dashed line indicates $T_{\text{ex}} = 17.5$\,K where $f(J, T_{\text{ex}}, X_{\text{mol}})$ has a minimum value, $f_{\text{min}}(J, T_{\text{ex}}=17.5\ \text{K}, X_{\text{mol}})$. \textit{Right y-axis:} Nomarlized $f(J, T_{\text{ex}}, X_{\text{mol}})$ against the $f_{\text{min}}(J, T_{\text{ex}}=17.5\ \text{K}, X_{\text{mol}})$  for the $^{12}$CO $J = 2-1$ transition line. The gray dashed line repersents the value of $f_{\text{min}}(J, T_{\text{ex}}=17.5\ \text{K}, X_{\text{mol}})$.
\label{fig:functionofTex}
}
\end{figure}

To model outflows, \citet{dunham2010} chose values of $T_{\text{ex}}$ ranging between 10 and 100 K. In particular, they used $T_{\text{ex}}=17.6$~K to calculate lower limits of the outflow properties. We follow a similar approach, finding the minimum value of $f(J, T_{\text{ex}}, X_{\text{mol}})$ and the corresponding excitation temperature $T_{\text{ex}}$ for the $^{12}$CO $J = 2-1$ transition line. Figure \ref{fig:functionofTex} shows the value of the $f(J, T_{\text{ex}}, X_{\text{mol}})$ versus $T_{\text{ex}}$ for the $^{12}$CO $J = 2-1$ transition line with the minimum value of the $f(J, T_{\text{ex}}, X_{\text{mol}})$ when $T_{\text{ex}}$ is 17.5 K. The right vertical axis of Figure \ref{fig:functionofTex} shows the enhancement factor of ejection mass to the minimum mass as a function of $T_{\text{ex}}$. For instance, the mass of ejected gas with $T_{\text{ex}}=100$ K is 2.5 times larger than the derived minimum mass with $T_{\text{ex}}=17.5$ K.\par

Since the high-velocity $^{12}$CO components were induced by the recent accretion activity, the $T_{\text{ex}}$ could be higher than 17.5 K. Additionally, as mentioned above, the presence of warm gas in the outflow indicates that the value of $T_{\text{ex}}$ might also greatly exceed the adopted value.
Nevertheless, We adopt $T_{\text{ex}}$ = 17.5 K to obtain the {\it minimum} value for the ejection mass. 

The ejection mass for each pixel and each velocity channel is calculated as 
\begin{equation}
M_{\textit{v}, \text{pixel}}=\mu_{\text{H}_{2}}m_{\text{H}}A_{\text{pixel}}N(\text{H}_{2})_{\textit{v}, \text{pixel}},
\end{equation}
where $\mu_{\text{H}_{2}}$ is the mean molecular weight per hydrogen molecule, with $\mu_{\text{H}_{2}}$ = 2.809 \citep{Evans2022}, $m_{\text{H}}$ is the mass of a hydrogen atom, and $A_{\text{pixel}}$ is the area of a pixel. The total H$_{2}$ column density, $N(\text{H}_{2})$, and total mass of ejected gas, $M_{\text{ejection}}$, are calculated by summing $N(\text{H}_{2})_{\textit{v}, \text{pixel}}$ and $M_{\textit{v}, \text{pixel}}$ over the pixels and velocity channels within a selected region and velocity range. To calculate the total column density and total ejection mass over the high-velocity ejection clumps, we select a 1\arcsec$\times$1\arcsec\ region corresponding to 1.0\arcsec\ to 0\arcsec\ for $\Delta\alpha$ and -0.5\arcsec\ to 0.5\arcsec\ for $\Delta\delta$ (the orange dashed boxes in Figure \ref{fig:b335_mom8_9_pv}). We also adopt the velocity range from -37.3 km s$^{-1}$ to -17.3 km s$^{-1}$. 
We use only the emission above 5$\sigma$ within the selected region and velocity range to obtain N$(\text{H}_{2})_{\textit{v}, \text{pixel}}$ and $M_{\textit{v}, \text{pixel}}$. Finally, the total H$_{2}$ column density and the total mass of the high-velocity ejection components are $2.1 \times 10^{23}$ cm$^{-2}$ and $7.5 \times 10^{-8} \; \text{M}_{\odot}$, respectively.

The kinetic energy, momentum, and force of ejected gas for each pixel and velocity channel are calculated following \citep{dunham2010}:
\begin{align}\label{eq:properties}
    \text{E}_{\textit{v},\text{pixel}} &= \frac{1}{2}M_{\textit{v},\text{pixel}} \times |\textit{v}_{\text{eject}}|^{2}, \nonumber\\
    \text{P}_{\textit{v},\text{pixel}} &= M_{\textit{v},\text{pixel}} \times |\textit{v}_{\text{eject}}|, \nonumber\\
    \text{F}_{\textit{v},\text{pixel}} &= M_{\textit{v},\text{pixel}} \times \frac{|\textit{v}_{\text{eject}}|}{\tau_{\textit{v}, \text{pixel}}}, \nonumber\\
    &\textit{v}_{\text{eject}} = \textit{v}-\textit{v}_{\text{sys}}
\end{align}
where $\tau_{\textit{v}, \text{pixel}}$ is the dynamical time for each pixel and each velocity channel. The value of $\tau_{\textit{v}, \text{pixel}}$ can be estimated by dividing the distance between the given position and the continuum peak position by the ejection velocity.

The total value of each ejection property (E$_{\text{ejection}}=\Sigma \, {\text{E}_{\textit{v}, \text{pixel}}}$, P$_{\text{ejection}}=\Sigma \, {\text{P}_{\textit{v}, \text{pixel}}}$, F$_{\text{ejection}}=\Sigma \, {\text{F}_{\textit{v}, \text{pixel}}}$) is derived by summing over the pixels and velocity channels. Following this approach, we estimated E$_{\text{ejection}}$, P$_{\text{ejection}}$, and F$_{\text{ejection}}$ to be 
1.9 $\times 10^{40}$\,erg, 
1.2 $\times 10^{-5} \; \text{M}_{\odot} \; \text{km} \; \text{s}^{-1}$, and 
3.8 $\times 10^{-6} \; \text{M}_{\odot} \; \text{km} \; \text{s}^{-1} \; \text{yr}^{-1}$, respectively. 

For the high-velocity $^{12}$CO outflow with the propagation velocity of $\sim$160 km s$^{-1}$ over the size of 1600 au by 1000 au, \citet{yen2010} reported F$_{\text{outflow}}=4.1 \times 10^{-5} \; \text{M}_{\odot} \; \text{km} \; \text{s}^{-1} \; \text{yr}^{-1}$ when scaled by the distance of 164.5 pc. This is approximately one order of magnitude higher than our result due to the larger coverage of the outflow. On the other hand, \citet{hirano1988} obtained F$_{\text{outflow}}= 1.8 \times 10^{-4} \; \text{M}_{\odot} \; \text{km} \; \text{s}^{-1} \; \text{yr}^{-1}$, when scaled by the distance of 164.5 pc, for the low-velocity $^{12}$CO outflow of B335. This low-velocity outflow is even more extended than the high-velocity $^{12}$CO outflow, making the outflow force almost two orders of magnitude higher than our result. 
In addition to the different scales, F$_{\text{outflow}}$ reported by previous studies must have been accumulated over a long period of time, while the F$_{\text{ejection}}$ calculated in this study is a value corresponding to one specific accretion event.

\subsection{Mass ratio between ejection and accretion} 
\label{sec:dis_rate_ejection_accretion}
In the early embedded stages, the central luminosity is mainly produced by accretion. Thus, the mass accretion rate, $\dot{M}_{\text{acc}}$, can be obtained from the following equation:
\begin{align}\label{eq:mass_acc}
    \dot{M}_{\text{acc}} &= \frac{L_{\text{acc}}\,R_{\star}}{f_{\text{acc}}\,GM_{\star}},
\end{align}
where $L_{\text{acc}}$ is the accretion luminosity, $R_{\star}$ is the radius of the protostar, $M_{\star}$ is the mass of the protostar, $G$ is the gravitational constant, and $f_{\text{acc}}$ is the fraction of the accretion energy that is radiated \citep{Hartmann2011}. 

We first estimate $L_{\text{acc}}$ to obtain the mass accretion rate during the timescale over which the high-velocity ejection clumps were ejected. We convert the W2 magnitude into W2 flux density (see right y-axis in Figure \ref{fig:b335_w2_lc}) as a proxy for the central luminosity.
The relation between the central luminosity and W2 flux density of B335 has been explored based on the 3D continuum modeling by \citet{Evans2023}; $L$/L$_{\odot}$ = [1.837+0.845($S_{\nu}$($W2$)$\times$0.67)+0.001($S_{\nu}$($W2$)$\times$0.67)$^{2}$].
The maximum luminosity, i.e., the outburst luminosity, is predicted by the model to be a factor of 5-7 higher than the quiescent luminosity of 3 \Lsun.
Alternatively, \citet{carlos2020} found a relation between the flux density of the mid-infrared and central luminosity for embedded protostars, $F_{\text{IR}}\;\propto L^{1.5}$. According to this relation, the outburst luminosity at MJD 58391 is enhanced by a factor of $\sim$5 over the quiescent luminosity at MJD 55304. This result is consistent with the luminosity calculated using the relation found in \citet{Evans2023}. 
Therefore, we adopt the relation between W2 flux density and central luminosity of B335 by \citet{Evans2023} to calculate the variation of central luminosity over the recent burst.

To calculate the mass accretion rate, we need some additional properties of B335 ($R_{\star}$, $M_{\star}$, and $f_{\text{acc}} = 0.5$).
In the 3D continuum modeling explored by \citet{Evans2023}, the luminosity depends upon $R_{\star}$ and the temperature, $T_{\star}$, of the source. 
\citet{Evans2023} held fixed $T_{\star}$ and varied $R_{\star}$ to match the observed quiescent luminosity, which is the SED modeling result using various photometric data observed before MJD 57000. The fixed value of $T_{\star}$ is 7000 K matches well the near-infrared part of the SED. 
To match the quiescent luminosity, $R_{\star}$ would be 1.17 $\text{R}_{\odot}$. We note that the near-infrared part of the SED may be contaminated due to various causes such as the rotation rate, inclination angle, outflow cavity properties, disk properties, stellar properties, and absorption of various ice molecules \citep{ylyang2017, Chu2020, Evans2023}. Thus, the fixed $T_{\star}$ of 7000 K has a large uncertainty as does $R_{\star}$. Nevertheless, here we adopt an $R_{\star}$ of 1.17 R$_{\odot}$ as the result of the best-fit model. 
\par
This same model provides an infall age, $t_{\text{col}} = 4 \times10^{4}$ yr, and mass infall rate, 6.26 $\times$10$^{-6}$ M$_{\odot}$ yr$^{-1}$ \citep{Evans2023}. Assuming all the mass that infalls from the envelope into the disk has fallen onto the star, $M_{\star}$ is 0.25 M$_{\odot}$. Other studies \citep{yen2015_b335, imai2019}, however, estimated the $M_{\star}$ based on the PV diagrams. \citet{yen2015_b335} derived the $M_{\star}$ of 0.05 M$_{\odot}$, while \citet{imai2019} provided a range of mass as 0.02 -- 0.06 M$_{\odot}$. Therefore, we adopt the $M_{\star}$ to range between 0.02 and 0.25 M$_{\odot}$.

\begin{figure}[htp]
\centering
{
\includegraphics[width=1\linewidth]{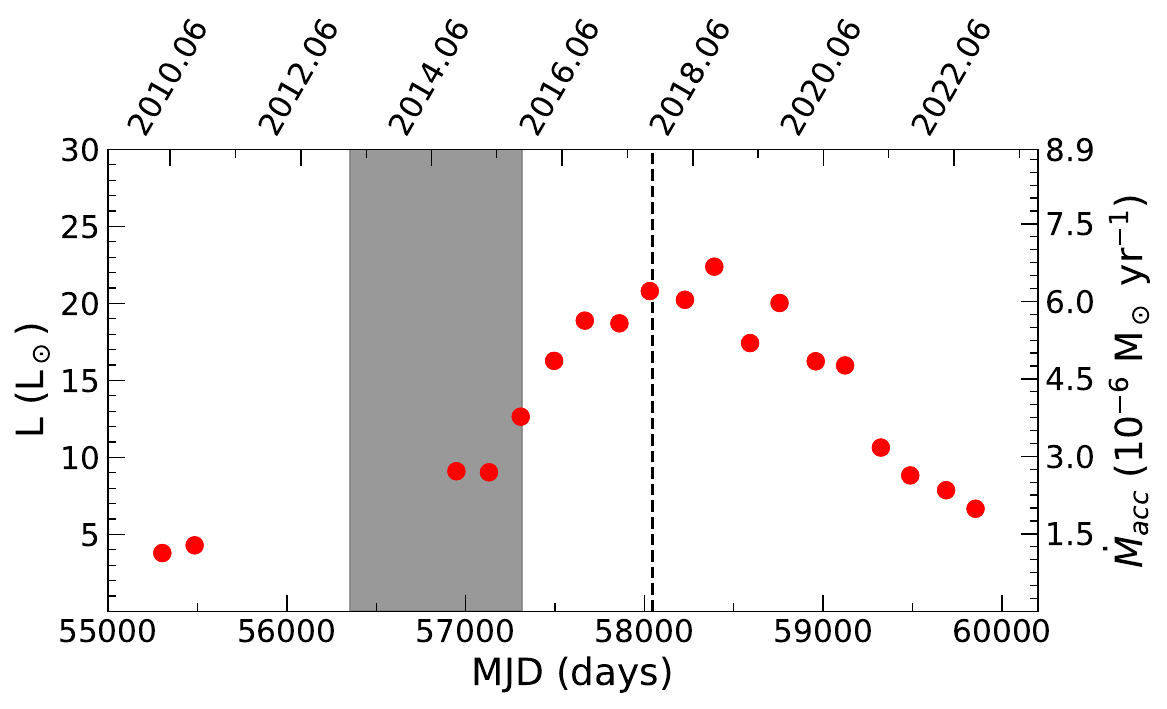}
}
\caption{Variation of luminosity and mass accretion rate of B335. The relation between luminosity and W2 flux density, $L$/L$_{\odot}$ = [1.837+0.845($S_{\nu}$($W2$)$\times$0.67)+0.001($S_{\nu}$($W2$)$\times$0.67)$^{2}$], was obtained by \citet{Evans2023}. Here $S_{\nu}$($W2$) is the W2 flux density. The gray-shaded region represents the timescale over which the high-velocity ejection components were ejected. The parameters used to estimate the $\dot{M}_{\text{acc}}$ in the figure are $M_{\star}$ of 0.25 M$_{\odot}$, $R_{\star}$ of 1.17 R$_{\odot}$, and $f_{acc}$ of 0.5.
\label{fig:b335_w2_lc_L_Mdot}
}
\end{figure}

$\dot{M}_{\text{acc}}$ is estimated using equation \ref{eq:mass_acc} and the values adopted above and presented in Figure \ref{fig:b335_w2_lc_L_Mdot}.
The average $\dot{M}_{\text{acc}}$ during the burst event (the gray shaded region, $\sim$2.6 yr) that launched the high-velocity ejection clumps is in the range of 
$\sim$3.1 $\times 10^{-6}$ to 
$\sim$3.8 $\times 10^{-5}$ M$_{\odot}$ yr$^{-1}$, depending on the adopted stellar mass. Therefore, the total mass accreted over the 2.6 yr, $M_{\text{acc}}$ becomes 
8.0 $\times\;10^{-6}$ to 1.0 $\times\;10^{-4}$ M$_{\odot}$.

The average mass ejection rate, $\dot{M}_{\text{ejection}}$, during the period when the high-velocity ejection components were ejected is estimated by dividing the total mass of the high-velocity ejection components by the timescale over which the ejection occurred, 2.6 yr,  resulting in 
2.9 $\times\; 10^{-8}$ M$_{\odot}$ yr$^{-1}$. Therefore, at least, $\sim$0.07$\%$ to $\sim$0.9$\%$ of the accreted mass was ejected by the recent burst event.

\section{Discussions}
\label{sec:uncertainty}
\subsection{Stellar properties} \label{subsec:stellar properties}
As discussed in Section 4.2, the estimation of $\dot{M}_{\text{acc}}$ depends on  stellar properties such as $L_{acc}$, $R_{\star}$, $M_{\star}$, and $f_{acc}$. We thus require accurate stellar properties to obtain an accurate mass accretion rate. The stellar properties adopted in this study are, however, quite uncertain.

First, for $L_{acc}$, the luminosity enhancement factor is uncertain at the 30\% level, ranging between 5 and 7 times the quiescent luminosity in \citet{Evans2023}. The relation, $L$/L$_{\odot}$ = [1.837+0.845($S_{\nu}$($W2$)$\times$0.67)+0.001($S_{\nu}$($W2$)$\times$0.67)$^{2}$],\
suggests the luminosity is enhanced by almost a factor of 7, and we adopt this value for our calculation. However, if the luminosity increases by only a factor of 5 over the quiescent luminosity, and the other stellar properties are fixed, then the derived $\dot{M}_{\text{acc}}$ would decrease by about 30\%.

Second, the best-fit model with $T_{\star} = 7000$\,K has $R_{\star} = 1.17$\,R$_{\odot}$. However, the fixed $T_{\star}$ of 7000 K is quite uncertain; thus, the 1.17 R$_{\odot}$ is poorly constrained. The typical protostellar radius is estimated as $\sim$3 R$_{\odot}$ \citep{Stahler1980}, which is 2.6 times larger. The $T_{\star}$ is 4370 K when we adopt the $R_{\star} = 3 $\,R$_{\odot}$. Using the typical protostellar radius and fixing all other parameters, the mass accretion rate would increase by a factor of 2.6.

Finally, $f_{acc}$ has a value as a $f_{acc}\;\leq\;1$ \citep{Hartmann2016}. Previous studies 
\citep{evans2015, yen2015_b335} have adopted a value of 0.5 and 1, respectively, for calculating the mass accretion rate of B335. This introduces an additional factor of 2 uncertainty in the mass accretion rate.

Therefore, the mass accretion rate varies with the adopted stellar properties, as discussed above. $\dot{M}_{\text{acc}}$ has a maximum value of 
$\sim$9.8 $\times\; 10^{-5}$ M$_{\odot}$ yr$^{-1}$ if the luminosity enhancement fator is 7, $M_{\star}$ = 0.02 M$_{\odot}$, and $R_{\star}$ = 3 R$_{\odot}$. On the contrary, $\dot{M}_{\text{acc}}$ has a minimum value of 
$\sim$2.2 $\times\; 10^{-6}$ M$_{\odot}$ yr$^{-1}$ if the luminosity increases by only a factor of 5, $M_{\star}$ = 0.25 M$_{\odot}$, and $R_{\star}$ = 1.17 R$_{\odot}$. The $f_{acc}$ is fixed at 0.5 in both cases. These extremes result in a factor of 45 difference in the derived mass accretion rate. Considering the uncertainties, the ejected mass ranges from 0.03$\%$ to 1.3$\%$ of the accreted mass during the recent burst event.

\subsection{Excitation temperature}
We adopt $T_{\text{ex}}$=17.5 K to determine the lower limit of the ejection mass. It is not unreasonable to adopt this value since the low-$J$ ($J = 2-1$) $^{12}$CO line is dominated by the coldest gas. However, as mentioned above, the $^{12}$CO $J=2-1$ line used in this study was induced by the recent outburst, and therefore $T_{\text{ex}}$ could be higher than the value we adopted.

If $T_{\text{ex}}$ is $\sim$200 K, the total ejection mass is a factor of 4.6 greater than when $T_{\text{ex}}$ is $\sim$17.5 K (See right y-axis in Figure \ref{fig:functionofTex}). The mass ratio of ejection to accretion also increases by the same factor. Thus, the ratio, $M_{\text{ejection}}/M_{\text{acc}}$ ranges from 0.03$\%$ to 6.0$\%$. To determine a more accurate mass ratio between accretion and ejection, other $^{12}$CO transition lines, such as $^{12}$CO $J = 3-2$, are required for an accurate calculation of $T_{\text{ex}}$.

\subsection{Inclination}
\label{subsec:Inclination}
The typical velocity of shock knots observed in outflows ranges 100 -- 150 km s$^{-1}$ \citep{bachiller1996, Dutta2020, cfLee2020, Jhan2022}. The velocity calculated with the inclination of 10$^\circ$ \citep{hirano1988} is consistent with the typical value. However, if we adopt the disk inclination of 87$^\circ$ from the line of sight, as derived by \citet{stutz2008}, which corresponds to the outflow inclination of 3$^\circ$ from the plane of the sky, the estimated velocity of knots 
reaches an extremely high velocity of $\sim$500 km s$^{-1}$.
In particular, the velocity and distance of the earliest ejected component are 483.1 km s$^{-1}$ and 142.4 au, while the velocity and distance of the latest ejected component are 549.8 km s$^{-1}$ and 70.1 au. If the velocity of the ejection components is related to the Keplerian velocity at the outflow launching radius, it can be calculated by $r_{0}$ = $GM_{\star}$/$\textit{v}_{\text{kep}}$$^{2}$. Thus, the launching radius is $\sim$0.02 R$_{\odot}$ for $M_{\star}$ = 0.02 M$_{\odot}$ and $\sim$0.2 R$_{\odot}$ for $M_{\star}$ = 0.25 M$_{\odot}$. This radius is smaller than the stellar radius we adopted. Nonetheless, adopting these velocities and distances, the high-velocity ejection components were ejected 1.4 to 0.6 years before the ALMA observation date. The $\dot{M}_{\text{ejection}}$ with outflow inclination of 3$^\circ$ would be 
9.5 $\times 10^{-8}$ M$_{\odot}$ yr$^{-1}$, which is calculated by dividing the $M_{\text{ejection}}$ by the timescale, 0.8 yr.\par
Then, the derived maximum and minimum $\dot{M}_{\text{acc}}$ (Equation~\ref{eq:mass_acc} and Section \ref{subsec:stellar properties}) during the period between 1.4 and 0.6 years before the ALMA observation date are 
$\sim$1.8 $\times 10^{-4}$ M$_{\odot}$ yr$^{-1}$ and 
$\sim$4.0 $\times 10^{-6}$ M$_{\odot}$ yr$^{-1}$, respectively. Thus, 0.05$\%$ $\sim$ 2.4$\%$ of the accreted mass must be ejected.
The ejection properties also slightly change with the small inclination. E$_{\text{ejection}}$, P$_{\text{ejection}}$, and F$_{\text{ejection}}$ are 
2.1 $\times\;10^{41}$ erg, 
4.0 $\times\;10^{-5}\;\text{M}_{\odot}\;\text{km}\;\text{s}^{-1}$, and 
4.2 $\times\;10^{-5}\;\text{M}_{\odot}\;\text{km}\;\text{s}^{-1}\;\text{yr}^{-1}$, respectively. 
\par
The inclination of the high-velocity outflow/jet is not necessarily the same as that of the larger slow outflow component. Therefore, it is important to know the correct inclination of the high-velocity ejection component to reduce the uncertainties associated with its physical parameters. For the accurate measurement of the inclination, monitoring observations of the high-velocity ejection component is critical.

\section{Conclusions} \label{sec:Conclusions}
 
We have revealed the contemporaneous accretion and ejection events that occurred recently in B335, using both ALMA and WISE/NEOWISE data.
From the PV diagram of the ALMA $^{12}$CO $J = 2-1$ image, a group of isolated high-velocity ejection components was revealed at $\sim-27.3$ km s$^{-1}$. The high-velocity CO gas components were likely ejected with the velocity of $\sim$156 km s$^{-1}$, about 4.6 to 2.0 years before the ALMA observation date. The WISE/NEOWISE light curve of B335 presents a brightening event by 2.5 mag at 4.6 $\mu$m over $\sim$8 years, with the maximum brightness in 2018.
A recent accretion event could have caused both the brightness increase of B335 and the ejection event, as recognized by the observed high-velocity ejection components.\par 
The ejection properties such as E$_{\text{ejection}}$, P$_{\text{ejection}}$, and F$_{\text{ejection}}$ for the high-velocity ejection components are estimated. To investigate the physical causality between contemporaneous ejection and accretion activities, we calculate the total mass of the high-velocity ejection components and the accreted mass for the same period. 
During the latest accretion burst, the mass ejection rate is 
$2.9 \times 10^{-8}\;\text{M}_{\odot}$ yr$^{-1}$, while 
the mass accretion rate ranges from 
$3.1 \times 10^{-6}\;\text{M}_{\odot}$ yr$^{-1}$ to 
$3.8 \times 10^{-5}\;\text{M}_{\odot}$ yr$^{-1}$, depending on the adopted physical parameters, in the timescale of about 2.6 years, suggesting that about 0.07 -- 0.9$\%$ of the accretion mass was ejected.\par
$\dot{M}_{\text{ejection}}$, $\dot{M}_{\text{acc}}$, the ratio between ejected mass and accreted mass, and ejection properties have uncertainty due to the uncertain parameters, particularly the 
inclination, adopted for the calculations. Therefore, constraining ejection physical parameters accurately is critical to understanding the physical causality between contemporaneous ejection activities and accretion events.

\section*{Acknowledgement}

This work was supported by the National Research Foundation of Korea (NRF) grant funded by the Korea government (MSIT) (grant number 2021R1A2C1011718). G.J.H.\ is supported by general grants 12173003 awarded by the National Natural Science Foundation of China.
D.J.\ is supported by NRC Canada and by an NSERC Discovery Grant. J.J.T. acknowledges support from  NSF AST-1814762.  
The National Radio Astronomy Observatory is a facility of the National Science Foundation operated under cooperative agreement by Associated Universities, Inc.

This paper makes use of the following ALMA data: ADS/JAO.ALMA\#2017.1.00288.
ALMA is a partnership of ESO (representing its member states), NSF (USA) and NINS (Japan), together with NRC (Canada), NSC and ASIAA (Taiwan), and KASI (Republic of Korea), in cooperation with the Republic of Chile. The Joint ALMA Observatory is operated by ESO, AUI/NRAO and NAOJ.
This publication also makes use of data products from NEOWISE, which is a project of the Jet Propulsion Laboratory/California Institute of Technology, funded by the Planetary Science Division of the National Aeronautics and Space Administration.

\bibliography{b335_ejection}{}
\bibliographystyle{aasjournal}

\end{document}